# Identifying Women with Mammographically-Occult Breast Cancer Leveraging GAN-Simulated Mammograms

Juhun Lee and Robert M. Nishikawa

***Abstract*— Our objective is to show the feasibility of using simulated mammograms to detect mammographically-occult (MO) cancer in women with dense breasts and a normal screening mammogram who could be triaged for additional screening with magnetic resonance imaging (MRI) or ultrasound. We developed a Conditional Generative Adversarial Network (CGAN) to simulate a mammogram with normal appearance using the opposite mammogram as the condition. We used a Convolutional Neural Network (CNN) trained on Radon Cumulative Distribution Transform (RCDT) processed mammograms to detect MO cancer. For training CGAN, we used screening mammograms of 1366 women. For MO cancer detection, we used screening mammograms of 333 women (97 MO cancer) with dense breasts. We simulated the right mammogram for normal controls and the cancer side for MO cancer cases. We created two RCDT images, one from a real mammogram pair and another from a real-simulated mammogram pair. We finetuned a VGG16 on resulting RCDT images to classify the women with MO cancer. We compared the classification performance of the CNN trained on fused RCDT images, $CNN_{Fused}$ to that of trained only on real RCDT images, $CNN_{Real}$, and to that of trained only on simulated RCDT images, $CNN_{Simulated}$. The test AUC for $CNN_{Fused}$ was 0.77 with a 95% confidence interval (95CI) of [0.71, 0.83], which was statistically better (p-value < 0.02) than the $CNN_{Real}$ AUC of 0.70 with a 95CI of [0.64, 0.77] and $CNN_{Simulated}$ AUC of 0.68 with a 95CI of [0.62, 0.75]. It showed that CGAN simulated mammograms can help MO cancer detection.***

## I. INTRODUCTION

Mammographically-occult (MO) cancer is a breast cancer that is visually occult, or very subtle, that radiologists failed to recognize. The detection of MO cancer is typically done by retrospective review, reviewing the negative prior mammogram using the diagnostic information (e.g., location) of the index year's mammogram. The incidence rate of MO cancer is high in women with dense breasts, as dense fibrograndular tissue could hide subtle signs of the presence of a breast cancer. The goal of our research is to triage women with dense breasts, who have a high risk of MO cancer, for additional screenings with magnetic resonance imaging (MRI) or ultrasounds, which have higher sensitivity than mammograms.

Our previous studies [1]–[3] showed that bilateral breast tissue difference is effective for detecting MO cancer. We showed that a novel image processing technique called Radon Cumulative Distribution Transform (RCDT) could amplify a very subtle MO cancer signal by exploiting lateral breast tissue differences. The RCDT technique for detecting MO cancer is based on the assumption that the existence of MO cancer in a mammogram can affect the left-right tissue symmetry, and therefore, MO cancer can be revealed by exploring the left-right breast tissue difference.

If we have a system to simulate a personalized mammogram with a normal appearance, we may be able to obtain additional information for detecting MO cancer by comparing simulated and real images. Mammography exams include the standard four views, left and right Cranial-Caudal (CC) views and left and right Mediolateral-Oblique (MLO) views. Each view can be used to detect any abnormal cues that may exist in the mammogram. In this respect, we may provide additional views by having simulated mammograms with a normal, healthy appearance for computerized methods and radiologists. However, simulating mammograms with a normal, healthy appearance is a difficult task. Simulating personalized breast mammograms is even more challenging.

Recently, machine learning and artificial intelligence communities have introduced a Generative Adversarial Network (GAN) to generate realistic *fake* photographs. A GAN consists of two convolutional networks, a generator and a discriminator. The generator tries to create a realistic *fake* image, which mimics the images in the database, while the discriminator tries to classify which one is the real image from the database, or the fake image from the generator. These two networks compete with each other during training. In the end, the generator learns the statistics of the given database and creates realistic/plausible images that the discriminator cannot tell if the images are the real or fake.

It is possible to train a GAN to simulate a plausible mammogram with a healthy, normal appearance. However, it is difficult to simulate individualized or personalized mammograms in this original form, as it is an unsupervised method; therefore, we cannot control the generation process.

A conditional GAN (CGAN) [4] is a type of GAN that solves image-to-image transition problems. Image-to-image transition

This work was supported in part by the National Institute of Health under Grant R37-CA248207.
Juhun Lee is with the Department of Radiology, Pittsburgh University, Pittsburgh, PA 15213 USA (email: leej15@upmc.edu).
Robert M. Nishikawa is with the Department of Radiology, Pittsburgh University, Pittsburgh, PA 15213 USA (email: RMN29@pitt.edu).

is a type of computer vision task that translates a given image to another in the target profile, which we can train from the given paired image dataset. Like other GANs, a CGAN consists of a generator and a discriminator. However, both the generator and the discriminator can observe what the given image is in the CGAN setup. Thus, the generator can use the information in the given input image to create the output, which enables a fully supervised setup for training the GAN. Specifically, the input image for the CGAN can guide the image simulation process by providing some key features that the generated image should have. For example, Isola et al. [4] used shoe sketches as input condition images to synthesize plausible shoe images. The sketches have key features, such as size, shape, and style of shoes.

The fact that women have two breasts with similar tissue profiles can help the CGAN to simulate personalized mammograms. Specifically, we can set one mammogram as the given or condition image to simulate the opposite side mammogram by providing key features for simulation, such as breast shape and density for the simulation. A given breast mammogram can serve as a sketch of the opposite breast for the CGAN, like the work of Isola et al. As the simulation is based on an individual's own breast, it can be a truly personalized simulated mammogram, which may help detect a possible MO cancer.

In this study, thus, we developed a CGAN that simulates the contralateral mammogram using a single-sided mammogram (either left or right) as input. We hypothesized that the simulated breast would exhibit a normal, healthy version of the contralateral breast, and we can obtain additional information for possible MO cancer by comparing the simulated and real breast mammograms.

We then used the RCDT as a primary image processing pipeline to analyze any given two mammograms. Two mammograms can be any combinations of the following four mammograms for each view (i.e., CC view or MLO view): 1) left, 2) right, 3) simulated-left, and 4) simulated-right. After the RCDT highlights suspicious signals for possible MO cancer, we used a Convolutional Neural Network (CNN) to analyze the resulting RCDT images to classify the cases with MO cancer out of the normal controls.

There are only a few previous studies investigating MO breast cancer among women with dense breasts. Mainprize and colleagues [5]–[9] focused on the masking effect of dense or fibro glandular tissue, instead of locating MO cancer directly. They used a model observer approach (pre-whitening model observer) to compute the probability of masking of a fixed size (5 *mm*) of a Gaussian shaped mass by dense breast tissue [5]. They estimated a breast density map using a commercial software (Volpara) on raw mammograms. Then, using the model observer and resulting density map, they estimated a detectability (1 – masking probability) map of a given mammogram. The resulting detectability map showed low intensity for the area that the human observer likely missed the lesion, while showing high intensity for the area that was easy for the human observer to find the lesion. Their recent work [9] utilized the above detectability map and volumetric breast density to stratify women with high masking risk who may benefit from supplemental screening. Specifically, they tried to find an operating point for deciding who needs to be sent for supplemental screening. They used 1897 cancer free women and 44 women with non-screen detected interval breast cancer. Compared to BI-RADS density and volumetric breast density alone, their method was able to find 64% of interval cancers with the least supplemental screenings.

Beyond the field of breast imaging, there exist previous works similar to our study, in terms of utilizing a GAN to identify abnormalities in given organs. For example, Schlegl et al., [10] used a GAN to detect abnormalities in the retina. They first trained a GAN on multiple patches from optical coherence tomography images of 270 normal retinas, to teach the network the distribution of a normal retina. Then, they used the trained GAN to detect abnormal lesions by mapping unseen images into a latent space spanned by a vector of random numbers that steer the image generation. Similarly, Alex et al., [11] used a GAN to identify lesions in brain MRI images. They also trained a GAN on multiple patches of non-lesion brain areas of 8 patients, to make it learn the underlying distribution of non-lesion brain areas. Then, they used a discriminator score of less than 0.5 as an indicator of brain lesions to detect. The most apparent difference between these previous works to our proposed study is on how we identified abnormal cases. While the previous studies used GANs as a main pipeline to detect abnormalities, we used a previously successful setup, RCDT, to detect MO cancer. As original GANs were not created for abnormality detection, they may not work as well as deep learning algorithms meant for detection.

We propose that left-right dense tissue differences are a key strategy to locate suspicious signals in mammograms that may indicate MO cancer [1]–[3]. We used the RCDT as the main image processing pipeline to amplify subtle left-right breast tissue differences that may pinpoint MO cancer. Specifically, by treating the left as a template and the right as a target, or vice versa, we can apply the RCDT to find what portion of the template should be changed or moved to make it look like the target. Thus, one may consider the RCDT as a projection algorithm, which projects a target in terms of a template. This projection provides the amplification of suspicious signals that may not be visible in a single breast mammogram. Using the RCDT to amplify the differences between left and right breast tissue is a key difference from the work of Mainprize et al. and success for pinpointing the location of MO cancer.

The current study is expanding our previous attempts [1]–[3] to detect MO cancer by adding simulated mammograms as supplemental information. As previously mentioned, one may treat the RCDT as a projection algorithm and having simulated mammograms can create another projection domain that may lead to improving MO cancer detection.

This paper is organized as follows. First, we developed a CGAN that can simulate a plausible mammogram of a given woman in section II. Then, we processed and combined the diagnostic information for detecting MO cancer from real and simulated mammograms by using the RCDT and deep CNN techniques in section III. We then evaluated the trained CNN to show the effectiveness of having CGAN generated mammograms for detecting MO cancer in section IV. We conducted post-hoc analysis on how a simulated mammogram could help detect MO cancer, compared to the other study, and discussed the limitations and areas of further improvements in section V.

## II. METHODS FOR SIMULATING MAMMOGRAMS

### A. Dataset

Under an approved institutional review board (IRB) protocol, we used a dataset that included screening full field digital mammograms (FFDMs) of 1366 women with normal/healthy breasts (BI-RADS category 1) and no prior breast surgeries from the University of Pittsburgh Medical Center (UPMC), to develop the CGAN for simulating the opposite side of the breast mammogram. The Selenia Dimension system (Hologic Inc, MA, USA) was used for all mammogram exams. All exams were acquired in 2018 and consisted of four standard views; left and right CC views and left and right MLO views. Each woman in this dataset had a screening mammogram at a single time point. Table I summarizes the characteristics of the mammogram dataset used for this study. BI-RADS breast density is based on the radiologist report.

TABLE I
Characteristics of Mammogram dataset for CGAN

| BI-RADS density level (5th edition) | # of samples |
|---|---|
| Entirely fatty | 290 |
| Scattered dense | 782 |
| Heterogeneously dense | 288 |
| Extremely dense | 6 |
| Age | # of samples |
| <40 | 12 |
| 40 to 50 | 259 |
| 50 to 60 | 418 |
| 60 to 70 | 425 |
| 70 to 80 | 209 |
| >80 | 42 |

### B. Preprocessing for simulating mammograms

We first converted the original 16-bit mammograms to 8-bit gray scale images. Using an existing algorithm [12] developed for breast density segmentation, we first segmented the breast area, dense tissue, and background automatically (Fig 1. b and e). We used only the breast area and removed any unnecessary portion (e.g., view-tag and non-breast tissue). After that, we resized each segmented image to the size of 1024 by 1024 pixels (Fig 1. c and f). CGAN requires two images, i.e., input and target, for training. We used left mammograms as the input and their corresponding right mammograms as the target. Fig 1 illustrates the above preprocessing steps.

### C. Conditional Generative Adversarial Network (CGAN)

We adopted the original CGAN setup, called pix2pix, by Isola et al. [4] for this study. The CGAN is trained to translate the given input image $x$ and random noise vector $z$ to the target image $y$, which can be formulated as $G: \{x, z\} \rightarrow y$, where $G$ indicates the generator. Generator $G$ is trained to fool discriminator $D$ by creating realistic fake images, while discriminator $D$ is trained to detect the images by the generator as fake.

The objective function of CGAN can be formulated as

$$Objective = \arg \min_G \max_D L_{cGAN}(G, D) + \lambda L_{L1}(G), \text{ Eq. (1)}$$

where $L_{cGAN}(G,D)$ and $L_{L1}(G)$ are the loss function for CGAN and $L_1$ regularization term, and written as

$$L_{cGAN}(G, D) = \mathbb{E}_{x,y}[\log D(x, y)] + \mathbb{E}_{x,z}\left[\log\left(1 - D(x, G(x, z))\right)\right], \text{ Eq. (2)}$$

$$L_{L1}(G) = \mathbb{E}_{x,y,z}[\|y - G(x, z)\|_1], \text{ Eq. (3)}$$

By having a normal mammogram of a woman as input image $x$ and the opposite mammogram of the same woman as target $y$, the above objective function set the generator $G$ to create a plausible normal mammogram that is similar to $y$. In other words, given a normal mammogram of a woman, the generator $G$ creates a normal opposite-side mammogram of the same women. Fig 2 illustrates how we adopted the original CGAN to simulate plausible mammograms.

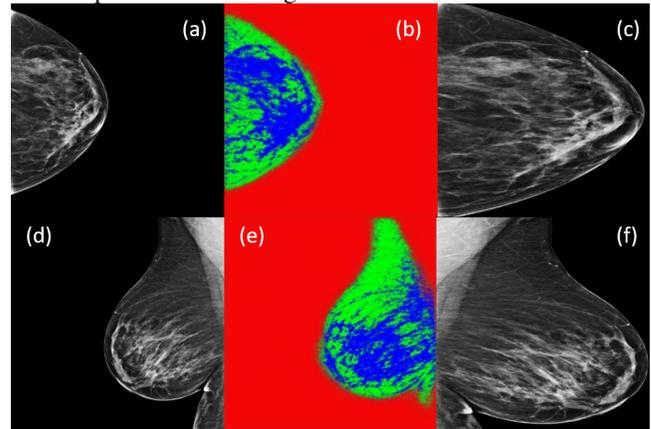

Fig. 1. This figure illustrates the preprocessing process for simulating mammograms. (a) and (d) are mammography images of the same women (Right CC and Left MLO view, respectively). (b) and (e) show the output images of the segmentation algorithm. The algorithm returns the segmentations of the breast area (green), dense tissue (blue), and background (red). (c) and (f) show the resulting images after locating the breast area, removing the non-breast area, segmenting the breast area with a tight rectangular window, and resizing it to have 1024 by 1024 pixels. We flipped left view mammograms (e.g., (f)) vertically so that all images have the same orientation. Note, we kept the pectoral muscle area, as it is required to generate plausible MLO view mammograms.

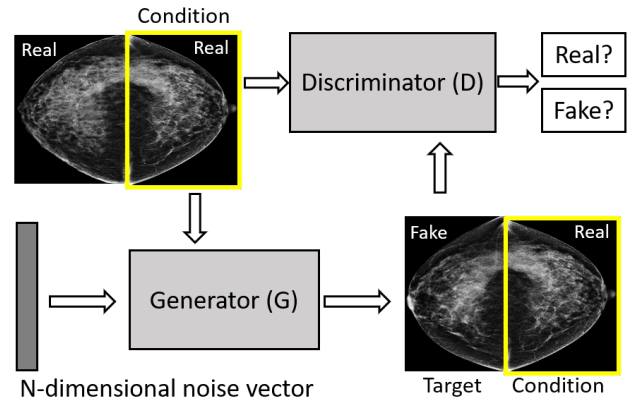

Fig. 2. This figure illustrates how we setup Generator (*G*) and Discriminator (*D*) for simulating plausible and normal mammograms. From N-dimensional noise vector and given condition normal mammogram of one woman, *G* and *D* are optimized adversativity to generate plausible and normal opposite sides of a mammogram of the same woman. Specifically, *G* takes real condition mammograms and noise vector to generate simulated mammograms, and *D* takes either real condition and target mammogram pairs or real condition and simulated mammogram pairs as input.

The generator in the CGAN uses U-net architecture as the skip connection between the encoder and the decoder can help to explore the similar characteristics that the input and the target images may have. This is also the right choice for our objective, as left-right breast mammograms should share common features (e.g., breast shape and density).

The original CGAN built to create 256 by 256 pixel images. It is too low in terms of resolution as the typical resolution of mammograms is 2k by 3k. Thus, we increased the depth of the generator by two levels (each depth doubles the output size) to create high resolution images of 1024 by 1024 pixels. One can use higher resolution, such as 2048 by 2028 pixels. However, to balance the simulation quality and computation burden, we used 1024 by 1024 pixels.

We directly adopted the discriminator from the CGAN, which is called *patchGAN*, where it focuses on the fidelity of $N$ by $N$-pixel patches, instead of evaluating the entire image. *PatchGAN* consists of four convolutional layers with 64, 128, 256, and 512 filters followed by 1D convolution and sigmoid function. As a result, the receptive fields of the discriminator $D$, i.e., *patchGAN*, was 70 by 70, where it only penalizes structure at the scale of 70 by 70-pixel patches.

### D. Training Details

We used the Adam optimizer [13] with a learning rate of 0.0002, and momentum parameters of $\beta_1=0.5$, $\beta_2=0.999$. In addition, we set the maximum epoch as 200 and the weight for $L_1$ regularization, $\lambda$, as 100, and the minibatch size of 1. We used random left-right vertical flip as data augmentation. We used a single Nvidia Titan X GPU with 12Gb memory for training. The training of the CGAN took approximately 48 hours.

We developed two CGANs, one for the CC-view and another for the MLO-view, as they look different (pectoral muscle is only visible in MLO view). In our previous study [14], we evaluated the performance of the CGAN for creating feasible contralateral mammograms by comparing the similarity between the simulated-real mammogram pairs (SR) to that of the real left-right mammogram pairs (RR). We used the mean squared error (MSE) and 2D correlation as surrogate measures for the similarity between two mammograms (either SR or RR mammogram pairs). We showed that the similarity of SR pairs is higher than that of RR pairs.

Fig 3 shows simulation outcomes for example CC and MLO view mammograms of one woman that was not used for training the CGAN.

## III. METHODS FOR DETECTING MAMMOGRAPHICALLY-OCCULT BREAST CANCER

### A. Dataset

Under an approved institutional review board (IRB) protocol, we used the screening FFDMs of 333 women with dense-breast tissue rated as BI-RADS breast density (4th edition) level 3 or level 4. All mammograms were acquired before 2014 using either Lorad Selenia or Selenia Dimension systems (Hologic Inc, MA, USA). Among the 333 women, 236 were normal with two consecutive negative screening FFDMs, and 97 had unilateral MO cancer. We used all four standard views, left and right CC views and left and right MLO views. Note that we used the most recent negative prior mammograms for both MO cancer cases and normal controls and therefore all 333 mammograms used for this study were normal (BI-RADS classification level 1). Thus, all cancer cases in this study showed no obvious visible signs of an abnormality in the mammogram. Table II shows the summary characteristics of the dataset used for this study.

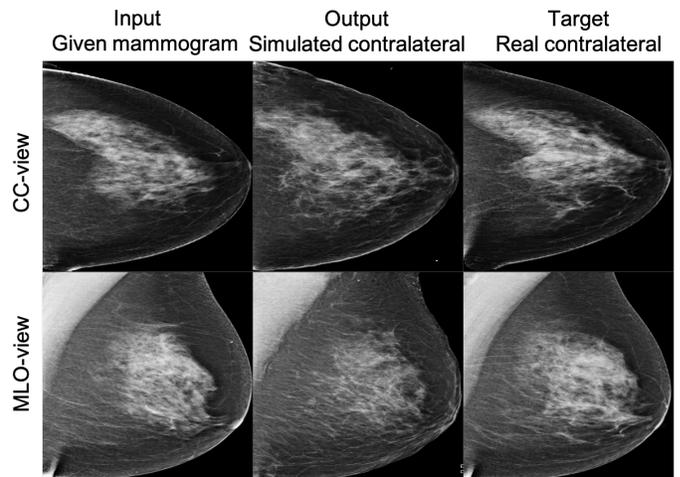

Fig. 3. This figure shows the simulated contralateral CC-view (first row, middle) and contralateral MLO-view (second row, middle) for an example normal woman, which was not used for training of the CGAN. The Generator in the trained CGAN is supposed to create a plausible normal breast mammogram of a woman using her mammogram as an input and a condition.

TABLE II
Characteristics of Mammogram dataset for MO cancer detection

|  |  | MO cancer | Controls |
|---|---|---|---|
|  |  | 97 | 236 |
| Age | Mean | 56.2 | 48.5 |
|  | [min, max] | [39, 70] | [36, 70] |
| BI-RADS breast density (4th edition) | 3 | 91 | 171 |
|  | 4 | 6 | 67 |
| Diagnosis[a] | IDC and DCIS | 40 | NA |
|  | IDC only | 27 |  |
|  | DCIS | 22 |  |
|  | IMC | 4 |  |
|  | ID and ILC | 2 |  |
|  | ILC only | 2 |  |

[a]IDC: Invasive Ductal Carcinoma, IMC: Invasive Mammary Carcinoma (a tumor that has mixed characteristics of both ductal and lobular carcinoma), ILC: Invasive Lobular Carcinoma, DCIS: Ductal Carcinoma In Situ.

### B. Simulating mammograms using trained CGAN

Similar to the previous section, we first segmented the breast area using the same algorithm [12], and cropped the segmented breast area using a tight rectangular box. The algorithm removed any soft tissue other than breast like Fig 1. Then, we resized each segmented image to the size of 1024 by 1024 pixels for CGAN.

For normal controls, we used the left mammograms as the input for the trained CGAN to simulate plausible normal versions of their corresponding right mammograms. For MO cancer cases, we used the normal side as the input to generate plausible normal versions of the contralateral breast mammograms, where the original target mammograms have MO cancer in it.

Note that we have two trained CGANs, one for the CC view and another for the MLO view. Using these CGANs, we simulated contralateral CC-view and MLO-view mammograms with the characteristics of normal, healthy breasts.

After the simulation, we had the following set of mammograms for each sample in the dataset as described in the list below:

*Normal controls*: 1) Left, 2) right, and 3) simulated right MLO, and 4) Left, 5) right, and 6) simulated right CC.

*MO cancer cases*: 1) Normal, 2) MO cancer side, and 3) simulated normal version of MO cancer side MLO, and 4) Normal, 5) MO cancer side, and 6) simulated normal version of MO cancer side CC.

### C. Radon Cumulative Distribution Transform (RCDT)

RCDT is a non-linear and invertible image transform that can represent a given image $I$ in terms of a given template image $I_0$ [15]. Let 2D Radon transform on the image $I$ and the template $I_0$ be $\hat{I} = R(I)$ and $\hat{I}_0 = R(I_0)$, respectively. Then, the RCDT($I \mid I_0$) = $\bar{I}$ is given as,

$$\bar{I}(\cdot,\theta) = (f(\cdot,\theta) - id)\sqrt{\hat{I}_0(\cdot,\theta)}, \text{ Eq. (4)}$$

and its inverse transform given as,

$$I = R^{-1}\left(det(J(g))(\hat{I}_0 \circ g)\right), where\ g(t,\theta) = \begin{bmatrix} f^{-1}(t,\theta) \\ \theta \end{bmatrix}, \text{ Eq. (5)}$$

where $J(.)$ is Jacobian and *id* refers to identify function i.e., $r(x) = x$, t and $\theta$ refer to the displacement and angle in the sinogram. In addition, $f(.,\theta)$ warps $\hat{I}(\cdot,\theta)$ to $\hat{I}_0(\cdot,\theta)$, where it satisfies

$$\int_{-\infty}^{f(t,\theta)} \hat{I}(u,\theta)\,du = \int_{-\infty}^{t} \hat{I}_0(u,\theta)du, for\ all\ \theta \in [0,\pi], \text{ Eq. (6)}$$

By using the RCDT, we can represent how the intensities of image $I$ and their locations differ from $I_0$ in terms of each angle $\theta$ and displacement $t$ from the origin in image space.

Figure 4 illustrates how we computed the RCDT image (c) from the template (b) to the target (a), what the resulting RCDT image represents in image space (d), and how it is different from the direct left-right difference image (e).

Note that RCDT($I \mid I_0$) is in Radon space and therefore, it is difficult to present the visual cues that led to MO cancer detection. Thus, we transformed back RCDT($I \mid I_0$) into the image domain by applying the inverse Radon transform.

Equation (4) can be divided into two parts, the first part with function $f$ and the second part with the template in radon space. The first part highlights where the major differences exist between the template and target images. By multiplying this highlighting part on the template image, one can pinpoint the area of the template that is most different from the target.

For optimal use of the RCDT for developing an algorithm to detect MO cancer, one needs to take the MO cancer side as the template, while the opposite side as the target. By doing so, we may pinpoint the potential area of MO cancer in the mammogram that has the actual MO cancer in it. For this reason, for MO cancer cases, we selected the mammogram with the MO cancer side as the template and others as the target.

In our previous study, we showed the effectiveness of RCDT processed mammograms over raw mammograms by showing CNN trained on RCDT processed mammograms statistically performed better than that on raw mammograms [3].

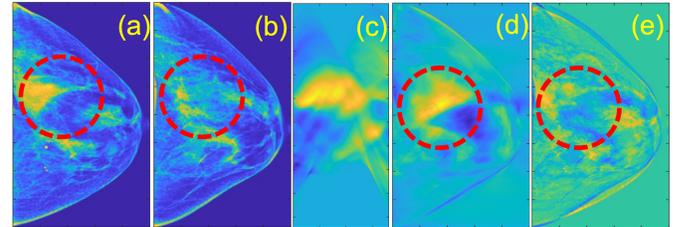

Fig. 4. Illustrates how we computed the RCDT image (c) from the template mammogram (b) to the target mammogram (a). (d) represents the resulting RCDT image in image space by taking the inverse Radon transformation. Compared to the direct left-right difference image (e), the RCDT image (d) highlights the left-right breast dense tissue difference shown in the dotted circle in (a) and (b), which failed to do so in the direct difference (e).

### D. Preprocessing for MO cancer detection

We resized all mammograms to have 800 by 500 pixels using bicubic interpolation. After that, we matched the orientation of the right and left mammograms by horizontally flipping the right-side mammograms.

There was a total of 6 mammograms per case, and therefore we can compute a total of 4 RCDT images, two from the CC-view, another two from the MLO-view. For normal controls, we set the left mammogram as the template, while we set the real right mammogram and simulated the left mammogram as the target, resulting in two RCDT images for each view. Likewise, for MO cancer cases, we can get two RCDT images for each view by setting the MO cancer side as the template and the real non-MO-cancer side and simulated MO cancer side as the target. After that, we enhanced the contrast of the resulting RCDT images by applying the contrast limited adaptive histogram equalization (CLAHE).

We then fused two RCDT images for each view by using the *imfuse* function, MATLAB (Mathworks, MA). The *Imfuse* function combines two images by assigning gray to the pixels with similar intensities, while setting different colors (e.g., magenta, green) for pixels with different intensities. This fusing method can provide additional information for detecting MO cancer, as they further highlight which area of the two RCDT images are different to each other. Fig 5 summarizes how we processed and combined real and simulated mammograms for training CNNs.

After that we resized the resulting 800 by 500 RCDT images to 400 by 250 for training CNNs. The reasons behind this choice are: 1) faster prototyping, 2) memory restriction, and 3) making the input size of typical CNNs pretrained on ImageNet [16]. The size of 400 by 250 pixels is larger than the input size of VGG16 [17] and ResNet [18] with the input size at 224 by 224, allowing room for data augmentation (e.g., random cropping, translation, etc).

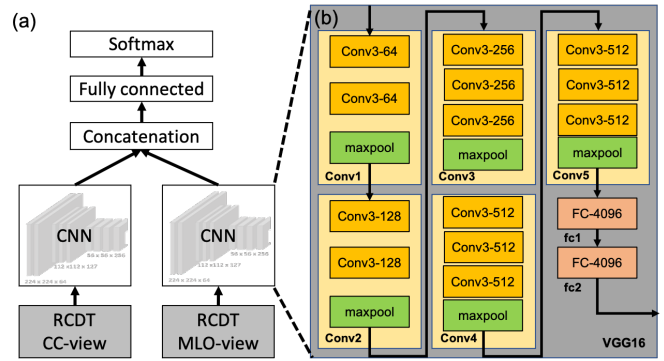

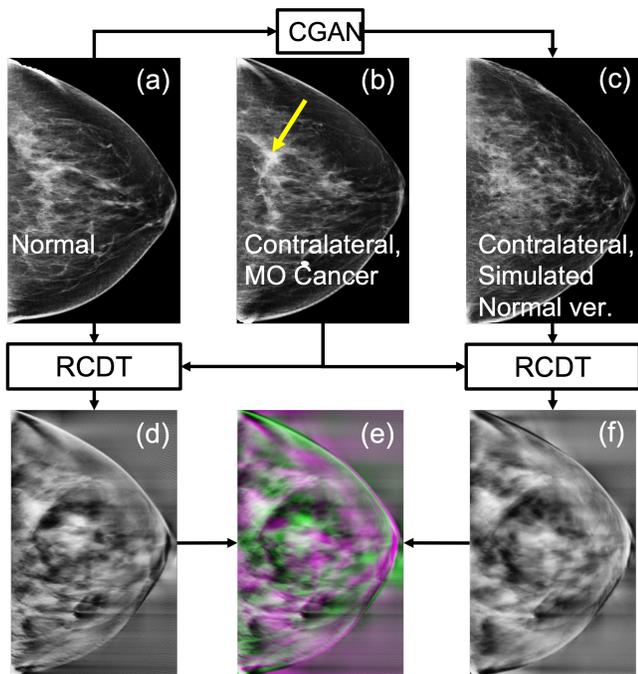

Fig. 5. This figure illustrates how we preprocessed mammograms for MO cancer detection. From normal side mammograms (a) of women with MO cancer, we generated a normal version (c) of the mammogram with the MO cancer (b) using the trained CGAN. The yellow arrow in (b) indicates the location of the MO cancer. We can then create two RCDT images, one (d) from (a) – (b) mammogram pair and another (f) from (b) – (c) mammogram pair. The RCDT images shown in this figure were contrast enhanced version. Then, we fused two RCDT images (e) to develop algorithms for MO cancer detection.

### E. CNN architecture for MO cancer detection

Figure 6.(a) shows the diagram of the CNN architecture that we used for this study. As CC and MLO views have different orientations and tissue presentations, we used two view-specific branch CNNs to analyze each view independently and then combined the processed information by each view-specific CNN at a high-level by using a concatenation layer, followed by a fully connected layer with two output and softmax layers to compute the case probability of the MO cancer.

The number of controls and MO cases are limited, therefore, we decided to utilize transfer-learning. We used CNNs pretrained on ImageNet and adopted, adjusted and fine-tuned those to solve our problem of interest.

The choice of CNNs in the proposed pipeline in Fig 6 can be diverse. One can use AlexNet[19], VGGNet[17], ResNet[18], DenseNet[20], and etc. Among those, we used VGG16, as their architecture is simple and showed compatible performance to state of art designs, like ResNet. We removed the last few layers of the pretrained VGG16, i.e., fully connected, softmax, and output layers, which were customed for classifying 1000 ImageNet categories. Fig 6. (b) shows the VGG16 network architecture after removing the last few layers that was used in the architecture shown in Fig 6. (a).

Fig. 6. (a) shows the CNN architecture that we used for analyzing multi-view RCDT processed mammograms. We used two CNNs, one for the CC-view and another for the MLO-view, as they show different imaging features. We then used a high-level fusion approach to combine the information analyzed by each view-specific CNN to compute the case probability of the MO cancer. (b) illustrates the VGG16 architecture that we used for this study. The last few layers for ImageNet classification was removed before inserted into the architecture shown in (a).

### F. Training Details

We divided our dataset into training, validation, and testing with a ratio of 7:1:2. Then, we used 5-fold cross-validation by assigning mutually exclusive 5 different portions of the entire dataset as the testing set in each cross-validation fold.

We implemented data augmentation including a horizontal and vertical random flip, random rotation with [-20°, 20°], random scale in the x-axis and the y-axis within the range of [0.5, 2], random translation within the range of [-20, 20] pixels and [-100, 100] pixels for the x-axis and the y-axis, respectively, and random cropping with the 224 by 224 window.

We used the NVIDIA Titan X graphic card with a memory of 12GB to fine-tune the above network. We set the learning rate for the weight and bias of the newly added layers to 10 times higher than the original layers to expedite the learning process for those new layers. That is, we set newly added layers in Fig 6 (a). with a learning rate of 10, while setting the learning rate of layers in Fig 6 (b) as 1. We used the Adam optimizer [13] and the cross-entropy loss to optimize the network. The training options included [max epoch: 128, mini-batch size: 16, learning rate: 0.00001, learning rate drop factor: 0.5, learning rate drop period: 10, validation patience: 5, random shuffling at each epoch]. We stopped the training early if there was no improvement in classifying MO cancer cases from controls.

### G. Evaluation

The purpose of this study was to show the improvement of MO cancer detection leveraging CGAN generated mammograms. Thus, we compared the CNN trained on RCDT images fused from real and simulated mammograms over that trained only on real mammograms. That is, we used the same CNN architecture shown in Fig. 6, but using different input images, one for CGAN leveraged RCDT images, another for RCDT images based only on real mammogram pairs, and another for RCDT images based only on real and simulated mammogram pairs. CGAN leveraged RCDT images will be (e) in Fig. 5, RCDT images based only on real mammogram pairs and real+simulated mammogram pairs will be (d) and (f) in Fig. 5, respectively.

The input size of VGG16 is 224 by 224 pixels and it cannot cover the RCDT image of 400 by 250 pixels. Thus, for evaluation, we used the sliding window approach with a window size of 224 by 224. We used the stride sizes of 5 and 35 for the x-axis and the y-axis, respectively. This yielded a total of 36 sliding window images (6 each for each axis) per case. To estimate the score for a case, we used the median value of the scores of the 36 sliding window images.

Then, we conducted ROC analysis and used the Area under the ROC curve (AUC) as our figure of merit.

Within the same cross-validation fold, we trained three composite CNNs, one using CGAN leveraged RCDT images or real and simulated fused, which we call $CNN_{Fused}$, another using RCDT images from real mammogram pairs, or $CNN_{Real}$, and another using RCDT images from real and simulated mammogram pairs, or $CNN_{Simulated}$, until they achieved appropriate validation performance (i.e., plateaued validation AUC), and then tested on the held-out test set.

Although we utilized the transfer learning technique, to be exact, fine-tuning of the ImageNet trained CNN for the given task, it is possible that having all weights updated every training iteration may not help to avoid local minima and over-fitting. In addition, our model has more weights to be trained, as there are two branch CNNs, one for the CC-view and another for the MLO-view, where each has millions of weights to be trained. This could be another risk factor for over-fitting. In fact, previous studies showed the merit of freezing lower level features for fine-tuning [21], [22]. Thus, we analyzed the effect of freezing layers. Starting from no freezing on all layers, we froze various levels of layers block from the first convolution block, Conv1, to the last fully connected layer, fc2, as shown in Fig 6 (b). We kept the same freezing scheme for two branch CNNs, i.e., if we froze up to Conv5 for the CC-view CNN branch, then we did the same for the MLO-view CNN branch.

## IV. RESULTS

For most cases, the training was completed when there was no improvement in the validation AUC, which was around 90 epochs. Then, we tested the trained CNN on the held-out test sets under 5-fold cross-validation. After that, we pooled the scores of the test sets of all 5-fold cross-validations and computed the ROC curve and the associated AUC value.

As the different levels of frozen layers could affect the classification performance, we first evaluated the effect of different levels of layer freezing. Fig 7 shows the pooled AUC (both validation and testing) for different levels of layer freezing. Note that the lower validation AUC was coming from the smaller number of cases (10% validation vs. 20% testing).

We found that freezing up to Conv5 yielded the best testing AUC. The case with no freezing (leftmost) resulted in the worst AUC. There is a clear trend, i.e., increasing in AUC from no freezing (leftmost) to freezing up to Conv5, and then the AUC value starts decreasing when we further freeze fully connected layers. The fact that no freezing yielded the lowest AUC supports our rationale for freezing layers; that is, the number of weights to be trained in the composite CNN in Fig 6 (a) is high and therefore, it is necessary to keep the relevant information of the CNN pretrained on ImageNet. As freezing layers up to Conv5 resulted in the best classification performance, we applied the same freezing procedures for the other two composite CNNs ($CNN_{Real}$ and $CNN_{Simulated}$) for comparison.

We repeated the pooled AUC approach as the above for the $CNN_{Real}$ and $CNN_{Simulated}$. Then, we used Delong's method [23] to compare the classification performances of $CNN_{Fused}$, $CNN_{Real}$, and $CNN_{Simulated}$. Specifically, we compared the AUC of $CNN_{Fused}$, 1) to that of $CNN_{Real}$ and 2) to that of $CNN_{Simulated}$.

Figure 8 shows the ROC curves of the three composite CNNs. $CNN_{Fused}$ showed an AUC of 0.77 with a 95% confidence interval (95% CI) of [0.71, 0.83], while that of $CNN_{Real}$ was 0.70 with a 95% CI of [0.64, 0.77] and $CNN_{Simulated}$ was 0.68 with a 95% CI of [0.62, 0.75]. The performance of $CNN_{Simulated}$ was compatible to that of $CNN_{Real}$. However, the performance of $CNN_{Fused}$ was statistically better (p-values < 0.02) than those of the $CNN_{Real}$ and $CNN_{Simulated}$. Specifically, the differences in AUC of $CNN_{Fused} - CNN_{Real}$ and $CNN_{Fused} - CNN_{Simulated}$ were 0.067 with a 95% CI of [0.011, 0.12] and 0.089 with a 95% CI of [0.032, 0.15]. The box plots of scores, as shown in Fig 9, for all CNNs confirms the above finding. From $CNN_{Simulated}$ to $CNN_{Fused}$, the score values for MO cancer cases increased. In addition, the class separation by $CNN_{Fused}$ was improved compared to those of $CNN_{Real}$ and $CNN_{Simulated}$.

These empirical results showed that we can get additional diagnostic information from simulated mammograms, when it was combined with the real mammograms.

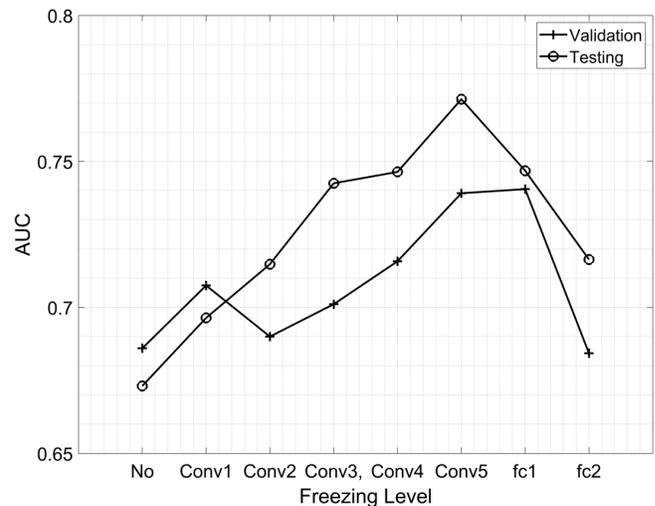

Fig. 7. Effect of freezing difference levels of layer blocks shown in Fig 6 (b). The AUC values clearly show that freezing up to Conv5 yielded the best AUC.

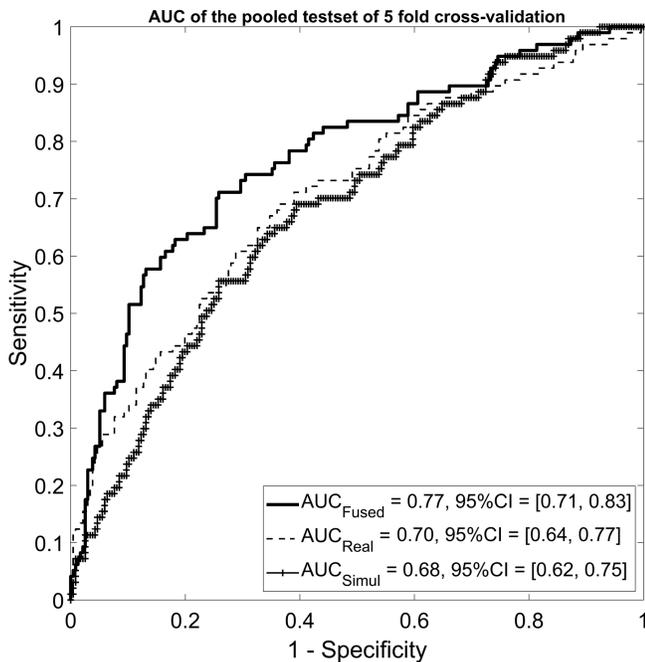

Fig. 8. This figure shows the ROC curves of three CNNs trained on fused CGAN leveraged RCDT images (solid line), RCDT images of real mammograms (dashed line), RCDT images of simulated mammograms (solid line with + marker). The CGAN leveraged CNN showed an AUC of 0.77, which was statistically better than that of other two CNNs.

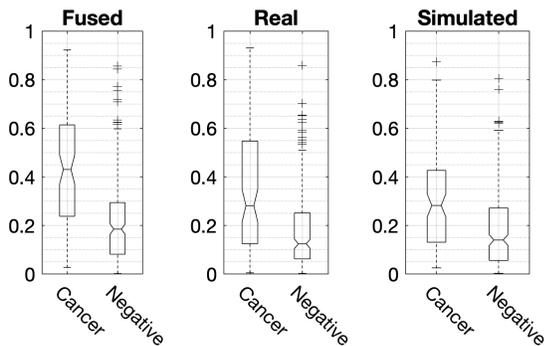

Fig. 9. This figure shows the box plots of MO cancer cases and normal controls for three composite CNNs on the pooled 5 independent held out test sets of 5-fold cross-validation. CNN using real and simulated fused showed the best class separation compared to those of the other two CNNs. In addition, the absolute scores for MO cancer were increased from simulated to real, and to fused cases. These results show that real and simulated RCDT images represent different visual cues for MO cancer such that they were a complement to each other for classifying MO cancer from normal controls. Each end point of box plots represents 25 and 75 percentiles of data. Central mark and notch are median and its 95% confidence intervals. + indicates outliers.

## V. DISCUSSION

In this study, we showed that the composite CNN trained on real and simulated fused RCDT images, $CNN_{Fused}$, performed better than that of the CNN trained only on real RCDT images, $CNN_{Real}$, as well as the case of the CNN trained on simulated RCDT images, $CNN_{Simulated}$. This proves our hypothesis, that is CGAN simulated mammograms can help MO cancer detection, empirically.

We further analyzed the effectiveness of having CGAN simulated mammograms for MO cancer detection. For this post-hoc analysis, we focused on two points: 1) whether simulated images degrade the cases that real images got correct and 2) how simulated images helped the cases that real images misclassified. To do so, we set 0.2 as the threshold to find the misclassified cases by the $CNN_{Real}$. We computed the threshold by taking the mid-point of two median scores, one from the MO cancer (0.28) and another from normal controls (0.12) for the $CNN_{Real}$ (second subfigure in Fig 9).

By setting a 0.2 score as the threshold, the $CNN_{Real}$ correctly classified 222 cases (63 MO cancers) and misclassified the remaining 111 cases (34 MO cancers). Fig. 10 (a) shows the box plots of all three CNNs for the samples that $CNN_{Real}$ got 100% correct. Although the overall $CNN_{Simulated}$ was lower than that of $CNN_{Real}$, $CNN_{Fused}$ was able to combine the information (real and simulated) to maintain the separation between the two samples. Fig. 10 (b) shows the box plots of all three CNNs for those of the 111 misclassified cases by the $CNN_{Real}$. The $CNN_{Simulated}$ was able to give better estimated scores for both MO cancer cases (i.e., higher scores than $CNN_{Real}$) and normal controls (i.e., lower scores than $CNN_{Real}$). By combining the information from the real and simulated RCDT images, the $CNN_{Fused}$ was able to give the better score (i.e., higher score than $CNN_{Simulated}$) for MO cancer cases, while it was not able to do so for normal controls. These results show that simulated mammograms did not degrade $CNN_{Fused}$, instead, simulated mammograms improved its performance by adding different views that real mammograms could not provide.

In addition to the post-hoc analysis on misclassified cases by the $CNN_{Real}$, we compared where and how each CNN focused on RCDT processed mammogram images. For this analysis, we applied Grad-CAM [24] on an example MO cancer case from misclassified cases by the $CNN_{Real}$. Grad-CAM highlights the area that leads to the CNN's decision. Specifically, it back propagates the estimated scores to the last convolutional block, in our case the maxpool in Conv5 in Fig 6 (b). By combining each channel's activation with back propagated weights, we can obtain the gradient guided class-activation map, i.e., grad-CAM for given images.

Figure 11 shows the input RCDT processed MLO-view mammograms for CNNs and their corresponding grad-CAM results. The yellow arrow indicates the MO cancer. For this example, the MO cancer was not visible in the CC view images and therefore omitted. The scores for $CNN_{Real}$, $CNN_{Simulated}$, and $CNN_{Fused}$ for the given example were 0.11, 0.22, and 0.63, respectively. The class-activation maps for $CNN_{Real}$ (Fig 11. g) and $CNN_{Simulated}$ (Fig 11. h) clearly show that the network missed the area with MO cancer, while that of the $CNN_{Fused}$ extended to the area with the MO cancer, which lead to a better estimation score for the MO cancer. This example shows the strength of combining the information from the real and simulated mammograms. The MO cancer (Fig 11. b) is visible, but very subtle even in the RCDT images from both the real and simulated mammograms (Fig 11. d and f). By combining these two RCDT images (Fig 11. d and e), we get an area of discordant between two RCDT images highlighted in magenta color in Fig 11. f (red dotted ellipsoid). This discordant area led the $CNN_{Fused}$ to extend its focused area to the real target, as indicated by the dotted red ellipsoid (Fig 11. i).

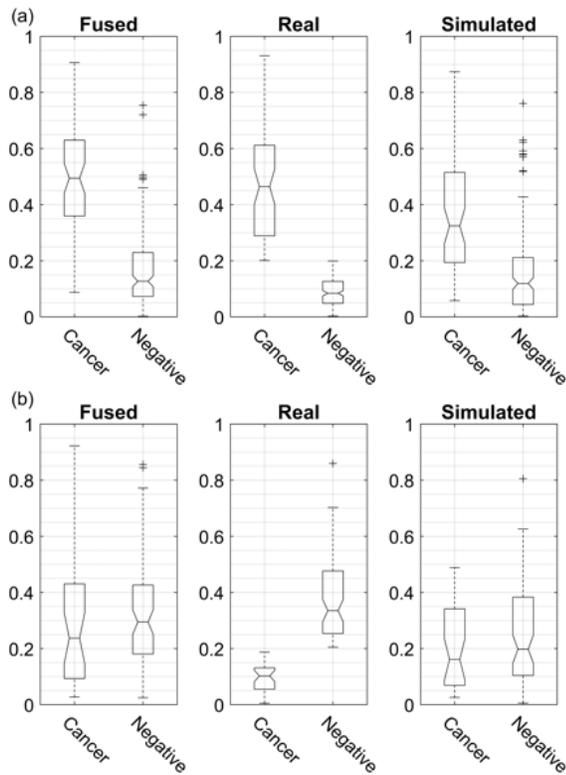

Fig. 10. This figure shows the box plots of MO cancer cases and normal controls for three composite CNNs, but for the cases that $CNN_{Real}$ was 100% correct (a) (N = 222, 63 MO cancers, 159 Normals) and misclassified (b) (N = 111, 34 MO cancers, 77 Normals). For (a), the scores of $CNN_{Simulated}$ for MO cancers are lower than that of $CNN_{Real}$ but it maintained the separation between MO cancer cases and controls. However, by combining the information from real mammograms and simulated mammograms, the scores for $CNN_{Fused}$ were higher than that of $CNN_{Real}$ for MO cancer cases, which made its separation between MO cancers and normal controls higher than that of $CNN_{Simulated}$. For (b), as shown in the box plots in the middle, $CNN_{Simulated}$ were able to estimate better scores for both MO cancer and control groups, which resulted in better estimation for $CNN_{Fused}$ for MO cancer cases.

We found the effectiveness of freezing layers for MO cancer analysis. Freezing up to the last convolutional blocks in Fig 6 (b). led to the best classification performance for analyzing RCDT images for MO cancer detection. Samala et al. [21] investigated the effect of freezing different levels of layer blocks for classifying malignant and benign breast lesions in Digital Breast Tomosynthesis (DBT). They used the AlexNet as a basis network and fine-tuned it on mammography images and then fine-tuned again on DBT images. Among various levels of freezing levels, they found that freezing up to the first convolutional layer blocks achieved the best classification performance, while not-freezing at all resulted in the worst classification performance. Their finding on the best way of freezing layers is different from ours, as we found the freezing up to the last convolutional layer block achieved the best classification performance on MO detection. The possible reasons for this difference could be: 1) the choice of the basis network, i.e., AlexNet vs. VGG16, 2) the imaging task, i.e., classifying benign-malignant lesions vs. MO cancer-normal controls, 3) the network complexity, i.e., single stream CNN vs. composite CNN with two branch CNNs, and 4) the use of different image processing techniques, i.e., patch-based vs.

RCDT processed from mammogram pairs. Further research would be required to investigate the reason for the different findings.

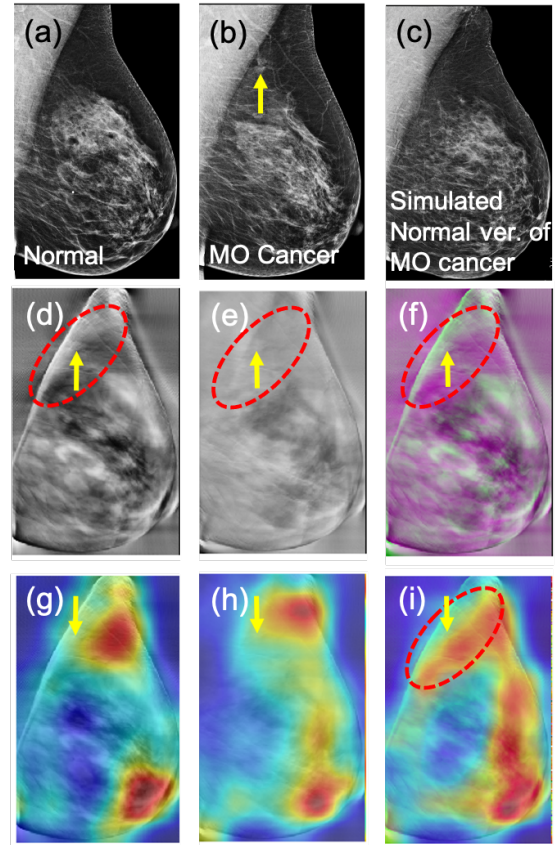

Fig. 11. This figure shows an example how a simulated mammogram can help the correct MO cancer classification. (a – c) are the normal side, the cancer side, and the simulated normal version of the cancer side of the MLO mammogram of a woman with MO cancer. Yellow arrow indicates the location of the MO cancer. (d and e) are RCDT images from the template (b) to the real target (a) and the simulated target (c), respectively. (f) is the fused RCDT images (d + e). (g), (h), and (i) are class-activation maps for (d), (e), and (f). The estimated scores for $CNN_{Real}$, $CNN_{Simul}$, and $CNN_{Fused}$ were 0.11, 0.22, and 0.63, respectively. MO cancer is visible, but subtle in RCDT images, and therefore $CNN_{Real}$ and $CNN_{Simul}$ missed the area with MO cancer as shown in (g) and (h). However, the area with MO cancer in two RCDTs are different, which resulted in a discordant area in the fused RCDT image. This led the $CNN_{Fused}$ to extend its focused area to the real target, as indicated by dotted red ellipsoid (i).

However, it should be noted that our finding on no freezing layers confirm that of Samala et al, which proves the importance of freezing layers for transfer learning. It has been a common problem for the field of medical imaging that having a limited number of available samples for training CNNs. Although there has been persistent effort on sharing imaging datasets in medicine (e.g., TCIA[25]), the majority of medical images is still private and not commonly shared with others. Therefore, transfer learning will still be one of the major applications for various medical image analysis tasks in the foreseeable future. Our finding (along with that of Samala et al.) suggests that research utilizing transfer learning should consider freezing layers to achieve the optimal target performance.

We observed artifacts in our CGAN simulated mammograms, specifically, checkerboard artifacts and artifacts at the nipple area (Fig 12). We performed post-hoc analysis to check if those artifacts affect the performance of $CNN_{Simulated}$. We randomly sampled 50 normal controls (approx. 22%) of the MO cancer detection dataset and found 11 cases with checkerboard artifacts and 7 with nipple artifacts. The mean and standard deviations of the $CNN_{Simulated}$ scores of the 11 normal cases with checkerboard artifacts were $0.236 \pm 0.227$ with [min, max] = [0.017, 0.761]. In the case of the nipple artifacts, the mean and standard deviations of the $CNN_{Simulated}$ scores were $0.069 \pm 0.069$ with [min, max] = [0.003, 0.213]. For most cases with checkerboard artifacts, the scores from $CNN_{Simulated}$ were lower than 0.2 (7 out of 11). For four cases with scores above 0.2, two cases were naturally difficult cases ($CNN_{Real}$ scores > 0.5). For the other two cases, $CNN_{Fused}$ was able to lower the final scores using information from $CNN_{Real}$ (scores < 0.1). For the cases with nipple artifacts, the scores were close to or lower than 0.2, showing the negligible effect of such artifacts on the performance of $CNN_{Simulated}$. These results indicate that $CNN_{Simulated}$ was able to learn the artifacts and remove them for MO cancer detection for most cases. For the few cases with $CNN_{Fused}$ > 0.2, the scores from $CNN_{Real}$ corrected the false information from the artifacts.

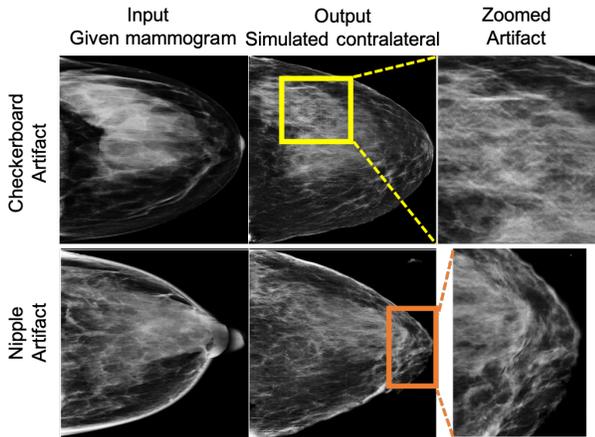

Fig. 12. This figure shows two examples with simulation artifacts, the first (top row) with the typical GAN-generated checkerboard artifact and another (bottom row) with a nipple artifact (simulated nipple as dense tissue). From the randomly sampled MO cancer dataset (N = 50), we found 11 checkerboard and 7 nipple artifacts.

There exist a few weaknesses of this study. We used a limited number of cases to train CGAN for simulating breast mammograms. The number of samples for training CGAN, i.e., 1366 women, is certainly not enough to cover all variations that mammograms could have. As one can see from Table I, there was also a limited number of women with extremely dense breasts. As our ultimate purpose is to find the women with dense breasts who have MO cancer, having more dense breasts (especially for extremely dense ones) is important. Having an insufficient number of dense breasts for training CGAN might have resulted in a negative impact on the performance of $CNN_{Fused}$ for MO cancer detection. However, even with such a limitation, we were able to show the effectiveness of using CGAN simulated mammograms for MO cancer detection. We expect that more data, especially more extremely dense breasts, for training CGAN will lead to better simulated mammograms, which will ultimately result in better performance for MO cancer detection.

Another weakness of this study is on the characteristics of the mammogram datasets. This is a single center and single vendor study. Our institute uses the Hologic system for breast cancer screening. As different vendors utilize different hardware and software configurations, e.g., image processing, detector, and x-ray spectra, breast tissue might appear different with different vendors. In addition, patient characteristics of one institute can be different from another. For these reasons, our method may not work on mammograms from other vendors and institutes. One may use fine-tuning to adapt the network on mammograms for other vendors and institutes. As future work, we will evaluate the performance of our method on mammograms from multi-vendors and multi-institutes and investigate fine-tuning for improving our method further.

In addition, we focused on identifying women with unilateral MO cancer only, which we believe is the first step and foundation for finding women with bilateral MO cancer. To do so, we simplified the inclusion of our dataset to have only women with unilateral MO cancer, which is the majority of women with breast cancer. We assumed that mammograms used as the condition image for the CGAN are normal, and those used as the template image for the RCDT are the MO cancer side, which can be otherwise in real clinical settings. For example, the CGAN synthesized image for the MO cancer side can exhibit artifacts, as it hasn't seen the cancer cases before. Moreover, RCDT may not be able to highlight the area with MO cancer if MO cancer is not present in the template image. These could degrade the performance of our proposed method. One way to solve this problem is to include additional mammograms to the pipeline. Specifically, we can create additional synthesize mammograms by taking each side (left or right) as condition images. Likewise, we can create additional RCDT processed mammograms by using each side (left or right) as template images. Then, we can add additional branch networks to the network, shown in Fig 6. (a), to accommodate newly generated images. We believe that such a modification can handle the cases with bilateral MO cancer. As future work, we plan to investigate the setup explained above.

There are some areas for improvement in this study. The first is on the choice of processing RCDT images. We used the magenta – green fusion to highlight the possible difference between two RCDT images. However, there are other ways to fuse given two RCDT images, which may improve MO detection performance further. One of the other fusing options would be assigning each RCDT image to different color channels, e.g., red and green. As we are using transfer learning, where CNNs pretrained on ImageNet use color images with three channels, by assigning RCDT images to each color channel, we may utilize the remaining channel for providing an additional image, e.g., original mammogram, which can further improve the $CNN_{Real}$'s performance on MO cancer detection. In addition, we can explore the patch-based approach (using patches of RCDT images with MO cancer and those of normal) to train CNNs, which can improve the MO cancer detection performance. Thus, we will investigate the optimal way of combining the diagnostic information from real mammogram pairs and real-simulated mammogram pairs as a future study.

The second area for improvement is on the choice of the basis CNN architecture for MO cancer detection. To build on our previous research [2], [3], we used VGG16. However, it is well-known that residual based networks, e.g., ResNet, typically perform better than the VGG16 network for various classification tasks. Nevertheless, the purpose of this study was to show the effectiveness of the additional diagnostic information provided by using CGAN generated mammograms, not developing the best working model. Future research, therefore, will include to find the best working basis network for MO cancer detection.

The third area for improvement is on how we combined two branch networks for estimating case-based probability of having MO cancer. We used the high-level fusion, i.e., having two networks to analyze the information in the CC and MLO views separately. We used the single fully-connected layer to process the concatenated information from two branch networks before the final classification (Fig 6. a). One can use multiple fully-connected layer blocks, like VGG16, to give more non-linearity before the final classification. It is also possible that the early fusion technique, sharing weights and gradients at the early level of layers, may help to locate the MO cancer. As there are many ways to combine information from each view, further research is necessary to find out which one is the best for MO cancer detection.

The fourth area of improvement is on improving the quality of the CGAN simulated mammogram. We found that the standalone performance of $CNN_{Simulated}$ was lower than $CNN_{Real}$, which could be improved by incorporating new or advanced techniques available for CGAN. For example, one can use interpolation-convolution for upsampling in Generator, instead of standard 2D transposed convolution, to remove checkerboard artifacts [26]. We will investigate new techniques that could improve our CGAN setup for synthesizing mammograms as a future study.

The last area of improvement is on investigating the possible extension of the proposed study for digital breast tomosynthesis (DBT). Many imaging centers in the US transitioned to mammography screening with DBT, as it provides increased cancer detection and fewer false positive findings [27]. Our proposed framework, that is utilizing CGAN, RCDT, and CNN for identifying MO cancer cases, can be extended. DBT, once reconstructed, is stacked 2D slice images of the breast. Thus, we may use our framework directly on slices of DBT volumes. Specifically, we can develop CGAN for simulating slices of DBT and then apply 2D RCDT and 2D CNN on real and simulated DBT slices. In addition, we can develop a CGAN that can simulate the entire 3D DBT volume by expanding the dimension of all components (e.g., layer dimension) of CGAN by 1. There also exists 3D RCDT (available at https://github.com/rohdelab/PyTransKit) and 3D CNNs (e.g., [28]), which allow our proposed technique/pipeline to be extended to identify MO cancer in DBT volumes. We will investigate the 3D version of our proposed framework on DBT images as a future study.

In summary and conclusion, we developed CGAN to generate a plausible mammogram with normal appearance using the opposite mammogram as a condition image, and empirically showed that CGAN generated mammograms can help detect MO cancer among women with dense breasts. Identified women can be triaged for additional screening with MRI, or ultrasound, which can result in earlier detection of MO cancer. Further research would require improving the quality of the generated mammograms, to find the optimal way to combine the generated mammograms with the real mammograms for MO cancer detection.

## ACKNOWLEDGMENT

The authors thank NIVIDIA for providing Titan X GPU for this research. Robert Nishikawa has research agreements with Hologic, Inc., Koios Medical, and GE Healthcare. He is on the advisory board of iCAD, Inc., and MaiData Corp.

## REFERENCES

[1] J. Lee, R. M. Nishikawa, and G. K. Rohde, "Detecting mammographically-occult cancer in women with dense breasts using Radon Cumulative Distribution Transform: a preliminary analysis," in *Medical Imaging 2018: Computer-Aided Diagnosis*, 2018, vol. 10575, p. 1057508.

[2] J. Lee and R. M. Nishikawa, "Detecting mammographically-occult cancer in women with dense breasts using deep convolutional neural network and Radon cumulative distribution transform," in *Medical Imaging 2019: Computer-Aided Diagnosis*, 2019, vol. 10950, p. 1095003.

[3] Juhun Lee and Robert M. Nishikawa, "Detecting mammographically occult cancer in women with dense breasts using deep convolutional neural network and Radon Cumulative Distribution Transform," *Journal of Medical Imaging*, vol. 6, no. 4, pp. 1–9, Dec. 2019.

[4] P. Isola, J. Zhu, T. Zhou, and A. A. Efros, "Image-to-Image Translation with Conditional Adversarial Networks," in *2017 IEEE Conference on Computer Vision and Pattern Recognition (CVPR)*, 2017, pp. 5967–5976.

[5] J. G. Mainprize, O. Alonzo-Proulx, R. A. Jong, and M. J. Yaffe, "Quantifying masking in clinical mammograms via local detectability of simulated lesions," *Medical Physics*, vol. 43, no. 3, pp. 1249–1258, Mar. 2016.

[6] J. G. Mainprize, O. Alonzo-Proulx, T. I. Alshafeiy, J. T. Patrie, J. A. Harvey, and M. J. Yaffe, "Prediction of Cancer Masking in Screening Mammography Using Density and Textural Features," *Academic Radiology*, vol. 26, no. 5, pp. 608–619, May 2019.

[7] J. G. Mainprize, O. Alonzo-Proulx, T. Alshafeiy, J. T. Patrie, J. A. Harvey, and M. J. Yaffe, "Masking risk predictors in screening mammography," in *14th International Workshop on Breast Imaging (IWBI 2018)*, 2018, vol. 10718, p. 107180E.

[8] O. Alonzo-Proulx, J. Mainprize, H. Hussein, R. Jong, and M. Yaffe, "Local Detectability Maps as a Tool for Predicting Masking Probability and Mammographic Performance," in *Breast Imaging*, 2016, pp. 219–225.

[9] O. Alonzo-Proulx, J. G. Mainprize, J. A. Harvey, and M. J. Yaffe, "Investigating the feasibility of stratified breast cancer screening using a masking risk predictor," *Breast Cancer Res*, vol. 21, no. 1, p. 91, Aug. 2019.

[10] T. Schlegl, P. Seeböck, S. M. Waldstein, U. Schmidt-Erfurth, and G. Langs, "Unsupervised Anomaly Detection with Generative Adversarial Networks to Guide Marker Discovery," in *Information Processing in Medical Imaging*, Cham, 2017, pp. 146–157.

[11] V. Alex, M. S. K. P, S. S. Chennamsetty, and G. Krishnamurthi, "Generative adversarial networks for brain lesion detection," in *Medical Imaging 2017: Image Processing*, 2017, vol. 10133, p. 101330G.

[12] Lee Juhun and Nishikawa Robert M., "Automated mammographic breast density estimation using a fully convolutional network," *Medical Physics*, vol. 45, no. 3, pp. 1178–1190, Feb. 2018.

[13] D. P. Kingma and J. Ba, "Adam: A Method for Stochastic Optimization," *arXiv:1412.6980 [cs]*, Dec. 2014.

[14] Juhun Lee and Robert M. Nishikawa, "Simulating breast mammogram using conditional generative adversarial network: application towards finding mammographically-occult cancer," presented at the Proc.SPIE, 2020, vol. 11314.

[15] S. Kolouri, S. R. Park, and G. K. Rohde, "The Radon Cumulative Distribution Transform and Its Application to Image Classification," *IEEE Transactions on Image Processing*, vol. 25, no. 2, pp. 920–934, Feb. 2016.

[16] O. Russakovsky, J. Deng, H. Su, J. Krause, S. Satheesh, S. Ma, Z. Huang, A. Karpathy, A. Khosla, M. Bernstein, A. C. Berg, and L. Fei-Fei, "ImageNet Large Scale Visual Recognition Challenge," *International Journal of Computer Vision*, vol. 115, no. 3, pp. 211–252, 2015.

[17] K. Simonyan and A. Zisserman, "Very Deep Convolutional Networks for Large-Scale Image Recognition," *arXiv:1409.1556 [cs]*, Sep. 2014.

[18] K. He, X. Zhang, S. Ren, and J. Sun, "Deep residual learning for image recognition," in *29th IEEE Conference on Computer Vision and Pattern Recognition (CVPR)*, Las Vegas, 2016.

[19] A. Krizhevsky, I. Sutskever, and G. E. Hinton, "Imagenet classification with deep convolutional neural networks," in *Advances in neural information processing systems (NIPS)*, 2012, pp. 1097–1105.

[20] G. Huang, Z. Liu, L. Van Der Maaten, and K. Q. Weinberger, "Densely connected convolutional networks," presented at the Proceedings of the IEEE conference on computer vision and pattern recognition, 2017, pp. 4700–4708.

[21] R. K. Samala, H.-P. Chan, L. Hadjiiski, M. A. Helvie, C. D. Richter, and K. H. Cha, "Breast Cancer Diagnosis in Digital Breast Tomosynthesis: Effects of Training Sample Size on Multi-Stage Transfer Learning Using Deep Neural Nets," *IEEE Transactions on Medical Imaging*, vol. 38, no. 3, pp. 686–696, Mar. 2019.

[22] N. Tajbakhsh, J. Y. Shin, S. R. Gurudu, R. T. Hurst, C. B. Kendall, M. B. Gotway, and J. Liang, "Convolutional Neural Networks for Medical Image Analysis: Full Training or Fine Tuning?," *IEEE Transactions on Medical Imaging*, vol. 35, no. 5, pp. 1299–1312, May 2016.

[23] E. R. DeLong, D. M. DeLong, and D. L. Clarke-Pearson, "Comparing the Areas under Two or More Correlated Receiver Operating Characteristic Curves: A Nonparametric Approach," *Biometrics*, vol. 44, no. 3, pp. 837–845, Sep. 1988.

[24] R. R. Selvaraju, M. Cogswell, A. Das, R. Vedantam, D. Parikh, and D. Batra, "Grad-CAM: Visual Explanations from Deep Networks via Gradient-Based Localization," in *2017 IEEE International Conference on Computer Vision (ICCV)*, 2017, pp. 618–626.

[25] K. Clark, B. Vendt, K. Smith, J. Freymann, J. Kirby, P. Koppel, S. Moore, S. Phillips, D. Maffitt, M. Pringle, L. Tarbox, and F. Prior, "The Cancer Imaging Archive (TCIA): Maintaining and Operating a Public Information Repository," *Journal of Digital Imaging*, vol. 26, no. 6, pp. 1045–1057, Dec. 2013.

[26] A. Odena, V. Dumoulin, and C. Olah, "Deconvolution and Checkerboard Artifacts," *Distill*, vol. 1, no. 10, p. e3, Oct. 2016.

[27] E. F. Conant, S. P. Zuckerman, E. S. McDonald, S. P. Weinstein, K. E. Korhonen, J. A. Birnbaum, J. D. Tobey, M. D. Schnall, and R. A. Hubbard, "Five Consecutive Years of Screening with Digital Breast Tomosynthesis: Outcomes by Screening Year and Round," *Radiology*, vol. 295, no. 2, pp. 285–293, Mar. 2020.

[28] D. Maturana and S. Scherer, "VoxNet: A 3D Convolutional Neural Network for real-time object recognition," in *2015 IEEE/RSJ International Conference on Intelligent Robots and Systems (IROS)*, 2015, pp. 922–928.